\pdfoutput=1
\documentclass[aps,prl,twocolumn,floatfix,showpacs,superscriptaddress,footinbib]{revtex4}
\usepackage{amsmath,amssymb,bm,url,color,verbatim,psfrag}
\usepackage[ansinew]{inputenc}
\usepackage{subfigure}

\ifx\pdftexversion\undefined
  \usepackage[dvips]{graphicx}
\else
  \usepackage[pdftex]{graphicx}
\fi

\def \be{\begin{equation}}
\def \ee{\end{equation}}
\def \bew{\begin{widetext}\begin{equation}}
\def \eew{\end{equation}\end{widetext}}
\def \bmlett{\begin{mathletters}}
\def \emlett{\end{mathletters}}

\def \omegar{\omega_R}
\def \omegam{\omega_M}
\def \xrms{x_{\rm zpt}}

\def \tA{\tilde{A}}
\def \tB{\tilde{B}}
\def \tchi{\tilde{\chi}}

\def \ra{\rightarrow}

\def \hx{\hat{x}}

\def \hF{\hat{F}}

\def \ha{\hat{a}}
\def \hb{\hat{b}}
\def \hc{\hat{c}}
\def \hH{\hat{H}}
\def \hd{\hat{d}}
\def \hf{\hat{f} }

\def \hxi{\hat{\xi}}
\def \heta{\hat{\eta}}

\begin{document}

\title{Quantum Noise Interference and Back-action Cooling in Cavity Nanomechanics}
%\title{Quantum noise interference as a route to ground state cooling in cavity electromechanics}

\author{Florian Elste}
\affiliation{Department of Physics, McGill University, Montreal, Quebec, Canada H3A 2T8}
\author{S. M.\ Girvin}
\affiliation{Department of Physics, Yale University, New Haven, Connecticut 06520, USA}
\author{A. A.\ Clerk}
\affiliation{Department of Physics, McGill University, Montreal, Quebec, Canada H3A 2T8}

\date{March 12, 2009}

\begin{abstract}
We present a theoretical analysis of a novel cavity
electromechanical system where a mechanical
resonator directly modulates the damping rate $\kappa$ of a
driven electromagnetic cavity. We show that via a
destructive interference of quantum noise, the driven cavity can
effectively act like a zero-temperature bath {\it irrespective}
of the ratio $\kappa / \omega_M$, where $\omega_M$ is the
mechanical frequency. This scheme thus allows one to cool the
mechanical resonator to its ground state without requiring the
cavity to be in the so-called `good cavity' limit $\kappa \ll
\omega_M$. 
%We also show that this system can be used to perform quantum-limited
%position measurements. 
The system described here could be
implemented directly using setups similar to those used in recent
experiments in cavity electromechanics.
\end{abstract}


\maketitle

\textit{Introduction.}---  Much of the rapid progress in fabricating and controlling nanomechanical devices
has been fueled by numerous promising technological applications.  
However, progress has also been motivated by the realization that such systems are ideally poised to allow the
exploration of several fundamental quantum mechanical effects.  Various studies have addressed
such issues as entanglement, quantum-limited measurement and Fock state detection in systems containing nanoscale mechanical
resonators~\cite{Armour02, Santamore04, Blencowe04, Clerk04c}. 
The issue of quantum back-action has also received considerable attention in these systems~\cite{Kippenberg08, Naik06}.
Understanding the back-action properties of a detector in a quantum electromechanical or optomechanical system is necessary if one wishes
to do quantum-limited position detection.  Further, this back-action can in some cases be exploited to achieve significant non-equilibrium cooling
of the mechanical resonator.  This is of particular importance, since a pre-requisite to seeing truly quantum behaviour in these systems is the ability
to cool the mechanical resonator to near its ground state.  

A particularly effective back-action cooling scheme is so-called 
``cavity cooling"~\cite{Dykman78, Braginsky02, Marquardt07, WilsonRae07}.  Here,
a mechanical resonator is dispersively coupled to a driven electromagnetic cavity 
(i.e.~the cavity's resonance frequency depends on the mechanical displacement $x$).  By monitoring the frequency of the cavity,
one can make a sensitive measurement of $x$.  In addition, the cavity photon number necessarily acts as a noisy force on the mechanics; 
for a suitably chosen cavity drive, this force can be used to effectively cool the mechanical resonator.  This approach has been used in a number of recent experiments, both with optical cavities coupled to mechanical resonators~\cite{Kippenberg07, Harris08}, as well as in
systems using microwave cavities~\cite{Regal08, Teufel08}.  Theoretically, it has been shown that such schemes could in principle allow ground state cooling of the mechanical resonator~\cite{Marquardt07, WilsonRae07}.  Similar physics even occurs in seemingly very different systems, e.g.~a superconducting single-electron transistor coupled to a mechanical resonator~\cite{Clerk05, Blencowe05}.

In this Letter, we present a theoretical analysis of a generic electromechanical (or 
optomechanical) system which is the dual of the dispersive coupling discussed above.  We again consider a 
driven cavity coupled to a mechanical resonator:  now, however, the mechanical displacement $x$ does not change
the cavity frequency, but rather changes its damping rate $\kappa$.  As we will discuss, such a coupling could be achieved
in a microwave cavity system, or in an optical cavity containing a moveable `mebrane-in-the-middle'~\cite{Harris08, Jayich08}.  
%Using the standard formalism of input-output theory, 
We
show that for such a dissipative cavity-mechanical resonator coupling, interference effects are important in determining the quantum back-action effects on the mechanical system;
this is {\emph not} the case for a purely dispersive coupling.
In particular,  we show that one can use destructive interference to allow the cavity to act as an effective zero-temperature bath, irrespective of
the ratio of the mechanical frequency $\omega_M$ to the cavity linewidth $\kappa$; as such, ground state cooling is possible without
requiring the `good cavity' limit $\omega_M \gg \kappa$.  This is in sharp contrast to the case of a dispersive
coupling, where ground state cooling is {\it only} possible if $\omega_M \gg \kappa$.  From a practical perspective, being able to deviate
from the good cavity limit is advantageous, as it allows one to use small drive detunings and hence achieve much
larger effective cavity-mechanical resonator couplings.  We show that this destructive interference effect persists in the case where one has both a dissipative and dispersive coupling; we also show that the dissipative coupling can also allow for a quantum-limited position measurement.

\textit{Model.}--- We consider a mechanical oscillator (frequency $\omegam$, displacement $x$) whose motion weakly modulates 
the damping rate $\kappa$ and resonant frequency $\omegar$ of a driven electromagnetic cavity.  For small displacements, 
both $\omega_R$ and $\kappa$ will have a linear dependence on $x$,
and we can describe the system via the Hamiltonian ($\hbar = 1$):
$ \hH =  \omega_R \ha^{\dagger} \ha +  \omega_M  \hc^{\dagger} \hc + 
		\sum_q  \omega_{q} \hb_{q}^{\dagger} \hb_{q} + \hH_{\rm damp} + \hH_{\rm int} + \hH_{\gamma} $.  
The first two terms describe the cavity and mechanical Hamiltonians, while $H_{\gamma}$ describes the intrinsic mechanical damping by an equilibrium
bath at temperature $T_{\rm eq}$.  The third term in $\hH$ describes the bosonic bath responsible for the dissipation and driving of the cavity.
Working in the usual Markovian limit where $\kappa \ll \omegar$, and where the bath density of states $\rho$ can be treated 
as energy-independent over relevant frequencies,  the cavity-bath interaction takes the form~\cite{Walls94, Clerk08b}:
\begin{eqnarray}
	\hH_{\rm damp} & = &
		-i  \sqrt{ \frac{\kappa}{2 \pi \rho} }  
		 \sum_q \left(  \ha^{\dagger} \hb_{q} -  \hb_{q}^{\dagger} \ha \right).
\end{eqnarray}
We stress that the Markovian limit is well justified both for cavities used in optomechanical and microwave electromechanical systems.  
The coupling between the cavity and mechanics is then described by:
\begin{eqnarray}
	\hH_{\rm int} & = & ( \hx / \xrms ) \left(
		\frac{1}{2} \tB \hH_{\rm damp} + \tA \kappa \ha^\dag \ha \right).
		\label{eq:Hint}
\end{eqnarray} 
The dimensionless coupling strengths $\tA,\tB$ above are given by
$ \tB \kappa =  ( d \kappa / dx) \xrms$, $\tA \kappa =  ( d \omegar / dx) \xrms$, where $\xrms$ is the zero-point motion amplitude of the mechanical oscillator.
Setting $\tB = 0$ recovers a purely dispersive coupling, while setting $\tA = 0$ corresponds to a purely dissipative optomechanical coupling.

\begin{figure}[tbp]
\begin{center}
\includegraphics[width=0.98\columnwidth]{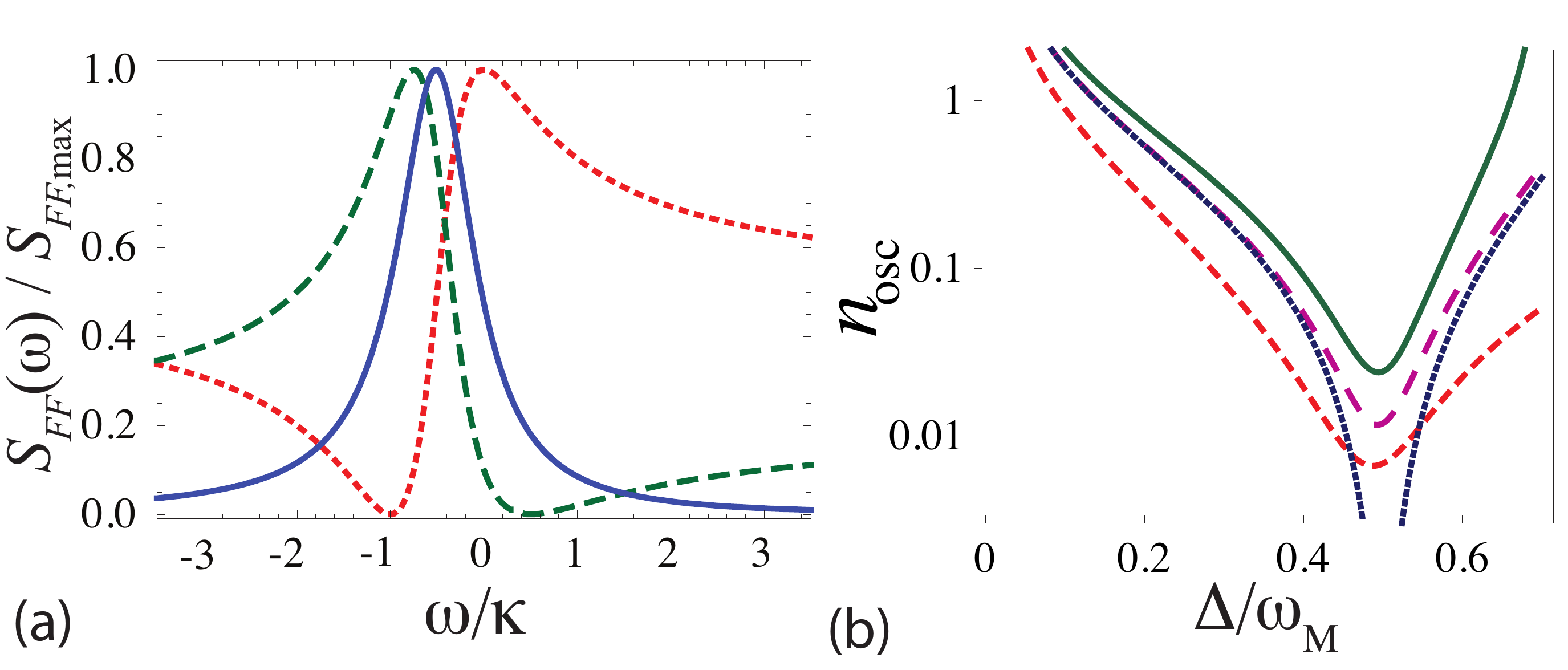}
\caption{(a) Back-action force noise spectral density for $\Delta = \kappa/2$,
for different ratios of the dispersive to dissipative optomechanical couplings.  
The solid blue curve corresponds to $\tA / \tB = 10$, the long-dashed green curve to
$\tA / \tB = 0.75$ and the short-dashed red curve to $\tA / \tB = 0$.  Each curve has been scaled
by its maximum value.  (b) Number of oscillator quanta $n_{\rm osc}$ versus drive detuning $\Delta$, obtained by 
solving the full equations of motion.
The dotted blue curve is the Bose-Einstein factor $n_{\rm eff}$ corresponding to the back-action effective
temperature $T_{\rm eff}$ when $\kappa = \omega_M$ (i.e. $n_{\rm eff} = S_{FF}(-\omega_M)/ (S_{FF}(\omega_M)-S_{FF}(-\omega_M))$.
The remaining curves correspond to $\tB \bar{a} = 0.2$,
$n_{\rm eq} = 50$, $\gamma = 10^{-6} \omega_M$, and 
$\kappa / \omega_M = 0.2$ (green solid curve),
$\kappa / \omega_M = 1.0$ (purple long-dashed curve),
$\kappa / \omega_M = 5.0 $ (red short-dashed curve).}
\label{fig:Fig1}
\end{center}
\vspace{-0.75 cm}
\end{figure}

\textit{Quantum noise.}---
We may now identify the quantum back-action force operator $\hat{F} \equiv  - (d/dx) \hH_{\rm int}$.  
For a sufficiently weak optomechanical coupling, linear-response theory applies, and the unperturbed quantum noise spectrum of $\hF$
%, calculated for $\tA=\tB=0$, 
will determine both the back-action damping and back-action heating 
of the mechanical resonator by the driven cavity \cite{Clerk05, Clerk08b}.  The relevant spectrum is  
$S_{FF}(\omega)=\int d\tau e^{i\omega\tau} \langle \hF(\tau) \hF(0) \rangle_0$.
Recall that $S_{FF}( \omega_M)$  is proportional to the Fermi's golden rule rate for the absorption of a mechanical quantum by the driven cavity, while
$S_{FF}( -\omega_M)$ is proportional to the corresponding emission rate.
The effective temperature associated with the back-action noise as seen by the mechanical oscillator is then given by
$k_B T_{\rm eff} \equiv  \omegam \left( \log \left[ S_{FF}(\omegam) / S_{FF}(-\omegam)  \right] \right)^{-1}$,
%\begin{eqnarray}
%	k_B T_{\rm eff} \equiv \hbar \omegam \left( \log \left[
%		\frac{ S_{FF}(\omegam) }{S_{FF}(-\omegam) } \right] \right)^{-1}
%\end{eqnarray}
while the back-action damping is given by $\gamma_{\rm BA} = \xrms^2 \left( S_{FF}(\omegam) - S_{FF}(-\omegam) \right)$.  These results
presume the total mechanical damping rate to be small enough that the resonator is only sensitive to noise at $\omega = \pm \omegam$.

We calculate $S_{FF}(\omega)$ by linearizing the full quantum dynamics about the solutions of the classical equations of motion (see below), and by making use
of the input-output formalism of quantum optics \cite{Walls94,Clerk08b}.  Letting $\Delta = \omega_{\rm drive} - \omegar$ be the detuning of the cavity drive, 
and $\bar a = \langle \ha \rangle$, one finds:
\begin{equation}
	\label{eq:SFF}
	S_{FF}(\omega) = \kappa  \left( \frac{  \tB |\bar{a}|}{2 \xrms}  \right)^2 
		\frac{ \left[ \omega+2\Delta - \frac{2 \tA}{ \tB}  \kappa  \right]^2}{(\omega+\Delta)^2+\kappa^2/4}.
\end{equation}

In the limit $\tB\ra 0$ of a purely dispersive coupling,  Eq.~(\ref{eq:SFF}) reduces to a simple Lorentzian, in agreement with Refs.~\cite{Marquardt07, WilsonRae07}.
This form has a simple interpretation:  it corresponds to the Lorentzian density of final states relevant to a Raman process where an incident drive photon
gains an energy $\hbar \omega$ while attempting to enter the cavity.  The optimal back-action cooling discussed in Refs.~\cite{Marquardt07, WilsonRae07} involves choosing 
$\Delta = - \omegam$ and $\kappa \ll \omegam$.  For these parameters, the drive photons are initially far from being on resonance with the cavity.  The absorption
of energy from the oscillator is resonantly enhanced, as it moves an incident drive photon onto the cavity resonance.  In contrast, emission of energy to the oscillator is greatly suppressed, as the drive photon is moved even further from the cavity resonance.  Thus, the $\tB=0$ form of $S_{FF}(\omega)$ and the resulting cooling are simply explained as a density of states effect.

In the more general case where the dissipative optomechanical coupling $\tB$ is also non-zero, we see that $S_{FF}(\omega)$ is {\emph not} a simple Lorentzian; as such, the back-action physics cannot be interpreted solely in terms of a density of states effect.  In general,  $S_{FF}(\omega)$ has an asymmetric Fano lineshape 
(see Fig.~\ref{fig:Fig1}); in particular, the cavity emission noise $S_{FF}(-\omega_M)$ is equal to zero {\it whenever} the detuning $\Delta = \omega_M / 2 + (\tA /  \tB) \kappa\equiv \Delta_{\rm opt}$.  Thus, for this value of the detuning, one finds that the cavity acts as an effective zero-temperature bath, irrespective of the ratio $\kappa / \omegam$.  For $\Delta = \Delta_{\rm opt}$, one has:
\begin{eqnarray}
	\label{eq:gammaBA}
	\gamma_{\rm BA, opt} = \tB^2   |\bar{a}|^2   \kappa 
		\frac{  \omega_M ^2}{\left( 3 \omega_M/2 + (\tA / \tB) \kappa \right)^2+\kappa^2/4}.
\end{eqnarray}

For $\Delta = \Delta_{\rm opt}$, the weak-coupling, quantum noise approach yields the equilibrium number of quanta in the oscillator 
to be $n_{\rm osc} = \gamma n_{\rm eq} / (\gamma_{\rm BA,opt } + \gamma)$, 
where $\gamma$ is the intrinsic damping rate of the oscillator, and $n_{\rm eq}$ is the Bose-Einstein factor associated with the bath temperature $T_{\rm eq}$.  One can thus cool to the ground state for a sufficiently large coupling and/or
cavity drive.  This possibility of ground state cooling for an arbitrary $\kappa / \omegam$ ratio is a main result of this paper.  Note that the optimal
detuning $\Delta_{\rm opt} $ is {\it greater} than $0$:  ground state cooling is possible even though the drive photons would seemingly need to ``burn off" energy to enter the cavity.  
This is in stark contrast to the purely dispersive case where a positive detuning leads to heating and a negative-damping instability; it highlights the fact that the back-action physics here is not simply a density of states effect.  Note finally that there is also an ``optical spring" effect associated with the back-action; it can be obtained directly from
$S_{FF}(\omega)$ \cite{Marquardt07}.  
%The fact that $\Delta_{\rm opt} > 0$ has the added benefit that one can avoid the static bistability associated with
%the dispersive coupling, as this occurs for negative detunings~\cite{Dorsel83}. 

A heuristic explanation of the Fano form of $S_{FF}(\omega)$ can be given.
Fano lineshapes arise generically as a result of interference between resonant and non-resonant processes;  the situation is no different here.  In the usual $\tB=0$ case, the only source of back-action
force noise is the number fluctuations of the cavity field $\ha$.  However, when $\tB \neq 0$, the mechanical oscillator mediates the coupling between the cavity and the cavity's dissipative bath.  As a result, it is subject to two sources of noise:  the shot noise associated with the driving of the cavity, as well as the fluctuations of $\ha$.  The first of these noise processes is white, 
whereas the second is not:  it is simply the shot noise incident on the cavity filtered by the $\omega$-dependent
cavity susceptibility.  The interference between these two noises yields a Fano lineshape for $S_{FF}(\omega)$, and the 
destructive interference at $\Delta = \Delta_{\rm opt}$ which causes $T_{\rm eff}=0$.  Note that Fano interference in quantum electromechanical
systems has recently been discussed in Ref.~\cite{Rodrigues09}, albeit in a completely different context.  

\textit{Equations of motion}.---
To address whether the destructive noise interference effect persists beyond the simplest weak-coupling regime, we
now examine the full solutions of the linearized Heisenberg equations of motion for our system.
% This will also allow us to address whether the destructive noise interference effect persists beyond the simplest weak-coupling regime.
For simplicity, we focus on the case $\tA = 0$ in what follows.  Working in frame rotating at the drive frequency, and writing $\ha = \bar{a}+\hd$, the linearized equations take the form:  
\begin{eqnarray}
	\dot{\hd}  & = & 
		\left( i \Delta - \frac{\kappa}{2} \right) \hd - \frac{\tB}{2} \kappa 
				\left( \bar{a} + \frac{\bar{b}_{\rm in} }{ \sqrt{\kappa}} \right) 
				\left(\hc + \hc^{\dag} \right)
			 - \sqrt{\kappa} \hxi, 
			 %- \sqrt{\kappa} \bar{b}_{\text{in}} A \hx  
		\label{eq:dEOM}
		\\
	\dot{\hc} 	& = & 
		-\left( i\omega_M+\frac{\gamma}{2} \right) \hc - \sqrt{\gamma} \heta 
		+ i \hf.
		\label{eq:cEOM}
\end{eqnarray}
Here $\bar{b}_{\rm in}$ is the amplitude of the coherent cavity drive, $\hxi$ describes vacuum noise entering the cavity from the drive port, and
$\heta$ describes thermal noise associated with the mechanical damping $\gamma$; both $\hxi$ and $\heta$ are operator-valued white noise~\cite{Walls94, Clerk08b}. 
The back-action force noise operator in Eq.~(\ref{eq:cEOM}) takes the form:
\begin{equation}
	\hf = i \frac{\tB}{2} \sqrt{\kappa} ( \bar{a}^* \hxi - \bar{b}_{\rm in}^* \hd) + h.c.
	\label{eq:littlef}
\end{equation}
We see the two contributions to the back-action discussed earlier:  the first term 
corresponds to the direct contribution of the drive shot noise, while the second term represents the contribution 
from fluctuations in the cavity field.  As discussed, the interference of these terms yields a Fano lineshape.

Upon solving the equations of motion, the oscillator spectrum $S_{cc} = \int dt \langle \hc^{\dag}(t) \hc(0) e^{-i \omega t} \rangle$ is found to be:
\begin{eqnarray}
	S_{cc}(\omega)  = 	|\tchi_M(-\omega)|^2 \left[ \sigma_{\rm eq}(-\omega) + S_{FF}(\omega)  \right]
	\label{eq:Scc}
\end{eqnarray}
where:
\begin{eqnarray}
	\tchi_M(\omega) & = & \chi_M(\omega) / \left[ 1 + i \chi_M(\omega) \Sigma(\omega) \right], 
		\label{eq:tchi} \\
	\Sigma(\omega) & = & 
		-i \tB^2 |\bar a|^2 \chi^*_M(-\omega) \chi_R(\omega) \chi^*_R(-\omega) \times
		\\
		&&
			 \omegam \Delta	
			\left[  \Delta^2 - \frac{3}{4} \kappa^2 +  i \omega \kappa \right],  
		\nonumber\\
	\sigma_{\rm eq}(\omega) & = &
		\gamma \left[
			n_{\rm eq} \left(1 + \frac{ \textrm{Re}  \Sigma(\omega) }{ \omega_M } \right)  	  +
		 \left| \frac{ \Sigma(\omega) }{ 2 \omega_M } \right|^2 (1 + 2 n_{\rm eq}) \right].  
		 \nonumber \\ \label{eq:sigma}
%	\sigma_{BA}(\omega) & = &
%		\left(  \tB / 2 \right)^2 \kappa \left|
%			\bar{a} + \sqrt{\kappa} \bar{b}_{\rm in} \chi_R^*(\omega)  \right|^2
%			\label{eq:sigmaBA}
\end{eqnarray}
Here, we denote the bare mechanical and cavity susceptibilties by $\chi_M(\omega) = (-i(\omega - \omega_M) + \gamma/2)^{-1}$
and $\chi_R(\omega) = (-i(\omega + \Delta) + \kappa/2)^{-1}$.  
Eq.~(\ref{eq:Scc}) has a simple interpretation:  the oscillator responds with a
modified susceptibility $\tchi_M$ to two independent fluctuating forces, corresponding to the two terms in the equation.  The first, described by $\sigma_{\rm eq}$, is the fluctuating thermal force 
associated with the intrinsic oscillator damping $\gamma$.  The second is the back-action from the driven cavity.
We see that its form is not affected by the coupling strength: the same spectral density $S_{FF}(\omega)$ (evaluated at $\tA=0$) found earlier
in the weak-coupling regime (c.f.~Eq.~(\ref{eq:SFF})) appears here.  Thus, the quantum noise interference discussed above continues to play
a role even for moderate coupling strengths.  A strong cavity-mechanical resonator coupling will nonetheless modify the physics, as 
the destructive interference occurring when $\Delta = \Delta_{\rm opt}$ only occurs at the single frequency $\omegam$.  
For a sufficiently strong coupling, the oscillator's total damping will become large enough that the oscillator will be sensitive to noise at 
frequencies away from $\omegam$, frequencies where the destructive interference is not complete.  As a result, the cavity will no longer appear to the oscillator as an effective zero-temperature bath. Note that when $\tA \neq 0$, the spectrum $S_{cc}(\omega)$ still has the general form given by Eqs.~(\ref{eq:Scc}, \ref{eq:tchi}, \ref{eq:sigma}):  one now simply
uses the full form of $S_{FF}(\omega)$, as well as a self-energy that contains extra terms $\propto \tA$.

\begin{figure}[!t]
\begin{center}
\includegraphics[width = 0.98 \columnwidth, angle=0]{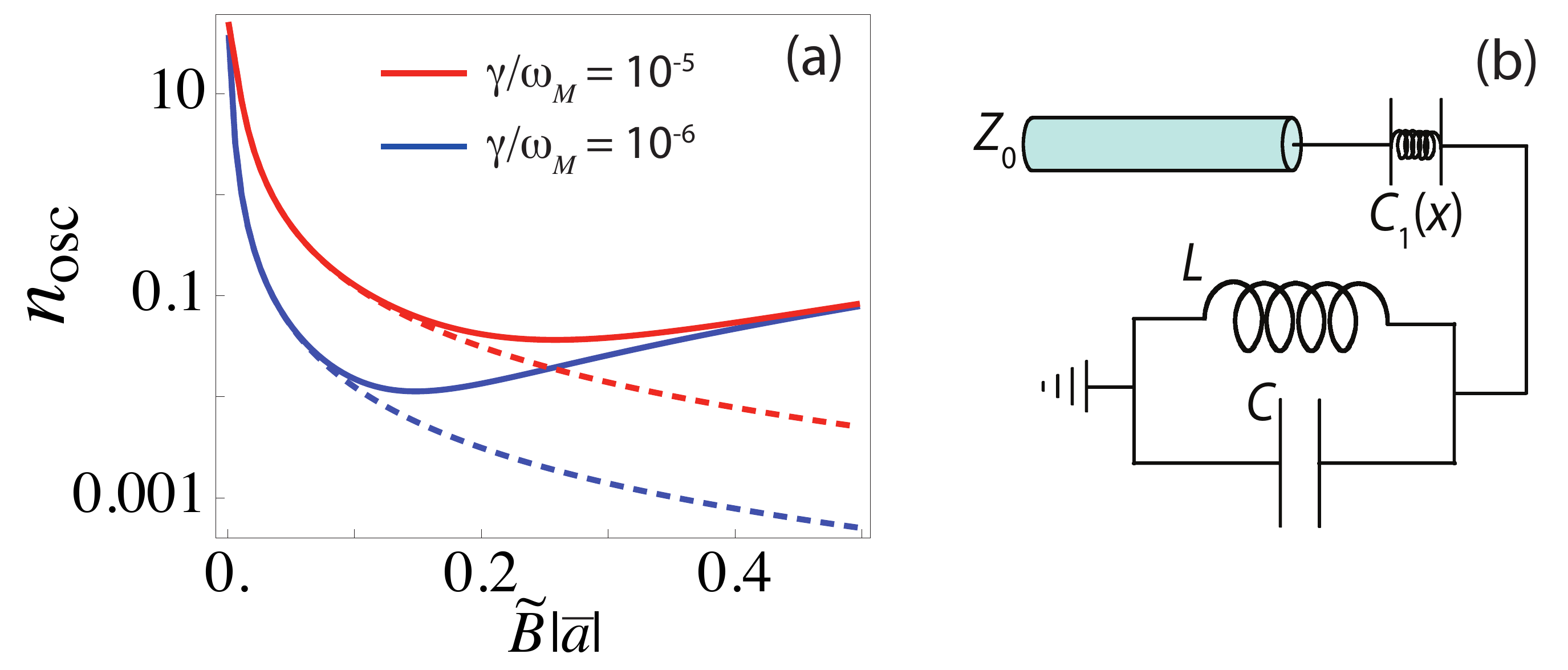}
\caption{(a) Number of mechanical quanta $n_{\rm{osc}}$ versus coupling strength
$\tB | \bar{a} |$, for $\tA=0$, $n_{\rm eq} = 50$, $\kappa / \omega_M = 1$ and for an optimal detuning $\Delta = \omegam/2$;
the mechanical damping $\gamma$ is as marked.
Dashed curves are results from the linear-response, quantum noise approach, whereas solid curves are obtained
from the full solutions of the equations of motion.  (b) Schematic of a cavity (modeled as an $LC$ resonator) coupled
to a transmission line (impedance $Z_0$) via an $x$-dependent capacitance $C_1(x)$.}
\label{fig:Fig2}
\end{center}
\vspace{-0.75 cm}
\end{figure}

To see whether this resonance-broadening effect as well as other strong coupling effects preclude ground state cooling at the
optimal detuning $\Delta = \Delta_{\rm opt}$, we calculate the average number of oscillator quanta $n_{\rm osc}$ directly by integrating 
$S_{cc}(\omega)$ in Eq.~(\ref{eq:Scc}).  In Fig.~\ref{fig:Fig2} we show 
the expected cooling for realistic parameter values, as a function of the cavity drive strength.
Strong-coupling effects lead to an optimal drive strength, beyond which $n_{\rm osc}$ starts
to increase.  Nonetheless, the minimum value of $n_{\rm osc}$ can still be significantly less than $1$.
%While the broadening of the mechanical resonance with increasing coupling does lead to a departure from the predictions of the weak-coupling theory,
%a significant amount of cooling remains possible.  Fig. YY shows the cooing as a function of oscillator frequency, for a fixed drive power $| \bar{b}_{\rm in}|$. 
Again, we stress that the dissipative coupling discussed here is in many ways advantageous to a system with a 
purely dispersive optomechanical coupling, as one does not need to use a large detuning to achieve a low effective temperature.

\textit{Physical realization.}---  The dissipative optomechanical coupling analyzed here could be realized in a microwave electromechanical system
similar to those studied in Refs.~\cite{Regal08,Teufel08}.  In such systems, a capacitor $C_1$ couples the cavity to a transmission line (impedance $Z_0$)
which both drives and damps the cavity.  One would now need to make $C_{1}$ mechanically compliant (see Fig.~\ref{fig:Fig2}b).  In general,
such a setup will have both dissipative and dispersive optomechanical couplings (i.e. $\tA \neq 0, \tB \neq 0$).  A careful analysis shows that in the physically
relevant regime $C_1 \ll C$, one can have $\tA \simeq \tB$ if the cavity impedance $Z_{R} = \sqrt{L/C}$ is made slightly smaller than $Z_0 \simeq 50 \Omega$.
For example, taking experimentally achievable parameters $\omega_R = 2 \pi \times  10 \textrm{ GHz}$, $C = 3.2 \textrm{ pF}$ and $C_1 = 0.01 C$ results in 
$\kappa / \omega_R \simeq 10^{-3} $ and $|\tA / \tB| \simeq 2.5$.
Eq.~(\ref{eq:SFF}) then implies that ground-state cooling via destructive noise interference is possible if one uses a drive detuning $\Delta = \Delta_{\rm opt} =  \omegam/2 + 2.5 \kappa$.  We stress this conclusion is independent of the magnitude of $\omegam / \kappa$; in contrast, if one had a purely dispersive cavity - mechanical resonator coupling, ground state cooling is {\it only} possible if $\omegam \gg \kappa$.  A second setup where the dissipative optomechanical coupling could be realized is an optical Fabry-Perot cavity containing a thin, moveable dielectric slab.  If the cavity is constructed with mirrors having different reflectivities, the motion of the membrane will naturally modulate the damping rate of the optical modes.  The resulting value of $\tB$ can be easily derived
using the approach of Ref.~\cite{Jayich08}.

\textit{Position detection}.---
A straightforward calculation shows that the power spectrum of $\hb_{\rm out}$, the output field from the cavity, is modified in a simple way by the 
coupling to the oscillator:  the oscillator adds a term $S_{xx}(-\omega) S_{FF}(\omega)$ (where $\omega$ is measured from the drive frequency); this holds for 
both $\tA, \tB \neq 0$.  Here, $S_{xx}(\omega) = \int d\tau e^{i\omega\tau} \langle \hx(\tau) \hx(0) \rangle$.  One can thus use the peaks at $\omega = \pm \omegam$ in the output spectrum to infer the temperature of the oscillator.  Note that for an optimal detuning $\Delta = \Delta_{\rm opt}$, the destructive interference effect means that the peak at $\omega = \omega_M$ in the output spectrum will vanish.  This could serve as an interesting test of our predictions.

Finally, specializing to the case where $\omegam \ll \kappa$, one can also use the dissipative optomechanical coupling to make a quantum-limited, weak continuous position measurement of the oscillator.  One drives the cavity off resonance, and performs a homodyne detection of the appropriate quadrature of the output field from the cavity.  One finds from Eqs.~(\ref{eq:dEOM}),(\ref{eq:cEOM}) that for any non-zero detuning $\Delta$, the corresponding back-action noise is as small as is allowed by the uncertainty principle, meaning that one can reach the quantum limit on the total added noise \cite{Clerk08b}.

\textit{Conclusions.}---
We have considered a generic system where a mechanical resonator modulates the damping of a driven quantum electromagnetic resonator, and demonstrated
how Fano interference is important to the back-action physics.  In particular, one can have a perfect destructive interference which allows the back-action
to mimic a zero-temperature environment; as such, such a system should allow near ground-state cooling of the mechanics.

We thank J. Harris for useful comments.  This work was supported by NSERC, CIFAR and the NSF under grants  DMR-0653377 and DMR-0603369.
 
%\ACcomment{Mention something about static bistability limiting detuning?  Mention fermionic realizations?  Other strong coupling effects?}

\bibliographystyle{apsrev}
\bibliography{ACrefsMar09}

\end{document}